\documentclass[fontsize=11pt,a4paper,DIV=15]{scrartcl}

\usepackage[utf8]{inputenc}                    
\usepackage[T1]{fontenc}                       
\usepackage{amsmath}                           
\usepackage[english]{babel}                    
\usepackage{hyperref}                          
\usepackage[
    sorting=none,                              
    autocite=inline,                           
    date=year,                                 
    citestyle=numeric-comp,                    
    bibstyle=numeric,                          
    giveninits=true,                           
    maxbibnames=10,                            
    maxcitenames=3,                            
    url=false,                                 
    sortcites=false,                           
]{biblatex}
\usepackage{graphicx}                          
\usepackage{tikz}                              
\usepackage{xcolor}                            
\usepackage{microtype}                         
\usepackage[draft=false]{scrlayer-scrpage}     
\usepackage{cleveref}                          
\usepackage{wrapfig}                           
\usepackage{blindtext}                         
\usepackage{abstract}                          
\usepackage{siunitx}                           
\usepackage{placeins}                          

\usepackage{helvet}                            
\usepackage{stix}                              
\usepackage{bm}                                

\graphicspath{ {figs/} }
\addto{\captionsenglish}%
{}               

\DeclareNameAlias{sortname}{family-given}      
\DeclareNameAlias{default}{family-given}       
\newcommand{\keywords}[1]{\vspace{0.5cm}\noindent\textbf{Keywords:} {#1} }

\definecolor{section}{Hsb}{195, 1, 0.227}
\definecolor{title}{Hsb}{198, 1, 0.31}
\definecolor{links}{Hsb}{198, 1, 0.545}

\KOMAoptions{
  abstract=false,
  titlepage=false,
  twocolumn=true,
}
\addtokomafont{title}{\color{title}}
\addtokomafont{subtitle}{\color{darkgray}}
\addtokomafont{disposition}{\color{section}}
\addtokomafont{date}{\normalsize\color{darkgray}}
\addtokomafont{author}{\LARGE\itshape\color{section}}
\addtokomafont{publishers}{\large}
\addtokomafont{captionlabel}{\bfseries}
\defcaptionname{USenglish}{\figurename}{Fig.}
\hypersetup{colorlinks=true,allcolors=links,allbordercolors=links}
\setcapindent*{0.5em}  

\title{Elastic shakedown and roughness evolution in repeated elastic-plastic contact}
\author{%
  Lucas Frérot\thanks{lucas.frerot@imtek.uni-freiburg.de, https://orcid.org/0000-0002-4138-1052}, Lars Pastewka\thanks{https://orcid.org/0000-0001-8351-7336}\\
  \textcolor{gray}{\small Department of Microsystems Engineering, University of Freiburg,}\\[-.35cm]
  \textcolor{gray}{\small Georges-Köhler-Allee 103, 79110 Freiburg, Germany}}
\date{\today}

\newcommand{\calC}{\mathcal{C}}

\newcommand{\body}{\mathcal{B}}
\newcommand{\tens}[1]{{\bm{#1}}}
\renewcommand{\epsilon}{\varepsilon}

\newcommand{\changed}[1]{{{#1}}}

\begin{document}

\twocolumn[%

\maketitle
\begin{abstract}
  Surface roughness emerges naturally during mechanical removal of material, fracture,
  chemical deposition, plastic deformation,  indentation, and other
  processes. Here, we use continuum simulations to show how roughness which is neither Gaussian nor self-affine emerges from repeated elastic-plastic
  contact of \changed{rough and rigid surfaces} on a flat elastic-plastic substrate. Roughness profiles change with each contact cycle, but appear to approach a steady-state long before the substrate stops deforming plastically and has hence ``shaken-down'' elastically. We propose a simple dynamic
  collapse for the emerging power-spectral density, which shows that
  the multi-scale nature of the roughness is encoded in the first few indentations.
  In contrast to macroscopic roughness parameters, roughness at small scales and
  the skewness of the height distribution of the resulting
  roughness do not show a steady-state, with the latter vanishing asymptotically with contact cycle.

\end{abstract}

]


\section{Introduction}
Friction, wear, thermal transfer and other tribological phenomena depend on the roughness of surfaces in contact: Its presence
means that contact only occurs on a limited subset of the apparent area of
contact, the \emph{true contact area} $A$, whose magnitude and geometry are of prime
importance for tribology~\autocite{vakisModelingSimulationTribology2018}.
Understanding the origins of roughness, with its multiscale nature~\autocite{renardConstantDimensionalityFault2013,perssonNatureSurfaceRoughness2005}, and making predictions on the true
contact area are therefore objectives of the study of rough
interfaces between solids. They are also both intimately linked to material properties.

Early models for contact of rough surfaces assumed that deformation is purely plastic~\cite{binderWiderstandKontakten1912,Holm1938-dh,Holm1938-va,Holm1941-zc,holmElectricContactsTheory2000}.
Assuming a surface hardness $p_\mathrm{m}$, this means that for a normal force $F_\mathrm{n}$ the true area of contact is given by $A=F_\mathrm{n}/p_\mathrm{m}$, because the mean pressure within contacts equals $p_\mathrm{m}$.
However, researchers in the 1950s and 60s discussed the idea of an ``elastic shakedown''~\cite{johnsonReviewTheoryRolling1966}: It seemed unlikely, that a contact would continue to behave plastically during repeated loading.
Eventually, the contact should ``shake down'' to a point where it only responds elastically.
This eventually lead to the development of elastic contact models, such as the famous 1966 model of Greenwood and Williamson~\cite{greenwoodContactNominallyFlat1966}.
The idea of elastic shakedown is probably one of the oldest ideas in the \changed{rough} contact literature.
Here, we present numerical calculations of the contact of rough surfaces with an elastic-plastic substrate to gain insights into the deformation processes during shakedown.


It seems reasonable to assume that a signature of shakedown is hidden in surface roughness:
Self-affine surface roughness is typically associated with irreversible deformation, such as
fracture~\autocite{mandelbrotFractalCharacterFracture1984,ponsonStatisticalAspectsCrack2016}
or
wear~\autocite{renardConstantDimensionalityFault2013,candelaMinimumScaleGrooving2016,milaneseEmergenceSelfaffineSurfaces2019,pham-baCreationEvolutionRoughness2021,aghababaeiHowRoughnessEmerges2022,garcia-suarezRoughnessEvolutionInduced2023},
but plastic deformation can also contribute to the emergence of roughness~\autocite{thomsonEffectPlasticDeformation1980,hinkleEmergenceSmallscaleSelfaffine2020,nohringNonequilibriumPlasticRoughening2022}.
%
In this work, we focus on the roughness created through \changed{quasi-static} repeated indentation (\changed{i.e. without sliding}), \changed{at a small contact area ratio,}
of an initially flat elastic-plastic half-space with a number of rough, rigid, \changed{periodic, isotropic} counter-faces, \changed{each different}.
Our work stands apart from prior theoretical work on elastic-plastic contact~\autocite{almqvistDryElastoplasticContact2007,tiwariContactMechanicsSolids2019}, which has investigated the contact of a rough deformable half-space with a rigid but flat plane.
While those calculations show smoothening of asperities, we find that the \changed{initially} flat counterbody continuously roughens during our calculations.
However, the statistical properties of the deformed body differ significantly from the indenting surface: Profiles are non-Gaussian and do not appear to be self-affine, even if the indenting surface has ideal self-affine properties.
Even though simple macroscopic properties, such as the RMS height, appear to saturate at a steady-state value after a few contact cycles, our calculations reveal that surface topography and subsurface plastic deformation continues to evolve beyond this apparent ``shake down'' of the contact.


\section{Methods}
\subsection{Elastoplastic Contact}
\begin{figure*}[h]
  \centering
  \includegraphics{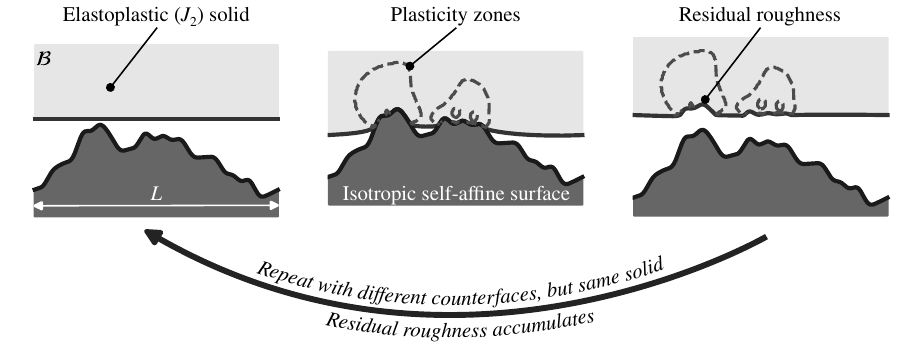}
  \caption{\changed{Two dimensional schematic view of the repeated indentation
      procedure we employ in this work. Indentation is made with a different
      surface at each iteration to reflect that in practice, contacts would be
      unlikely to occur twice with the same alignment of surfaces. $\body$ is a
      semi-infinite elastic-plastic solid obeying a $J_2$ isotropic linear
      hardening plasticity model. The rigid, rough, isotropic counterface has a period
      $L$.}}\label{fig:schematic}
\end{figure*}
\changed{\Cref{fig:schematic} illustrates the repeated indentation procedure described in the introduction. We simulate the contact between an elastic-plastic half-space $\body$ and a
periodic rigid rough surface using a volume-integral equation
approach~\autocite{bonnetImplicitBEMFormulations1996,jacqDevelopmentThreeDimensionalSemiAnalytical2002,frerotFourieracceleratedVolumeIntegral2019}.}
In these repeated contact simulations, \changed{without horizontal movement}, we denote $h_k$ the rigid surface of the $k$-th
indentation. 
Unlike conventional boundary-integral equation methods used in elastic and
elastic-plastic rough
contacts~\autocite{stanleyFFTBasedMethodRough1997,campanaElasticContactSelfaffine2008,yastrebovContactRepresentativeRough2012,almqvistDryElastoplasticContact2007,weberMolecularProbesReveal2018},
our approach fully considers plastic residual strains in the subsurface region of the half-space.
The subsurface (volumetric) plastic strain field contributes to 
displacements and stress~\autocite{frerotCrackNucleationAdhesive2020}.
Equilibrium equations are solved with Green's functions applied in the Fourier
domain~\autocite{frerotFourieracceleratedVolumeIntegral2019}.

The plasticity model assumes an additive strain decomposition and a von Mises ($J_2$) yield criterion, \changed{$f_\mathrm{y}$}:
\begin{align}
  \tens{\sigma} &= \calC:(\tens\epsilon - \tens\epsilon^p),\\
  f_\mathrm{y}(\tens\sigma) & = \sqrt{\frac{3}{2}}\|\tens{s}\|,\quad \text{with
                              } \tens{s} = \tens\sigma -
                              \frac{1}{3}\mathrm{tr}(\tens\sigma)\tens I.
\end{align}
We have defined $\tens \sigma$ the Cauchy stress tensor, $\tens \epsilon$ the total strain, $\calC$ the small-strain elasticity tensor, $\tens\epsilon^p$ the plastic strain and $\tens s$ the deviatoric stress.
The colon operator here introduces contraction on the two right-most indices of a tensor.
This type of model, also referred to as $J_2$ plasticity~\cite{simoComputationalInelasticity1998}, is the most widely-used canonical plasticity model.
It assumes plastic deformation progresses as volume-conserving laminar flow beyond a yield shear (von Mises) stress, but not under hydrostatic conditions.
The model is empirical and material-unspecific, but captures the basic phenomenology of plastic flow at macroscopic scales.

Since our simulations are quasi-static, and the pseudo-time variable is given by
the number of \changed{different} surfaces used in indentation, $k$, we describe plastic strain
rates as increments~\autocite{simoComputationalInelasticity1998}.
The equivalent cumulated plastic strain is defined as:
\begin{align}
  e^\mathrm{p} = \sqrt{\frac{2}{3}} \sum_{i=1}^k{\|\Delta{\tens\epsilon}^\mathrm{p}_i\|}.
\end{align}
The plasticity conditions can be written as, in terms of
$e^\mathrm{p}$ and $\sigma$:
\begin{align}
  f_\mathrm{y}(\tens\sigma) - f_\mathrm{h}(e^\mathrm{p}) & \leq 0,\\
  \Delta{e}^\mathrm{p} & \geq 0,\\
  \left(f_\mathrm{y}(\tens\sigma) -
  f_\mathrm{h}(e^\mathrm{p})\right)\Delta{e}^\mathrm{p} & = 0,
\end{align}
where $f_\mathrm{h}(e^\mathrm{p}) = \sigma_0 + E_\mathrm{h}e^\mathrm{p}$ is the linear
hardening function with initial yield stress $\sigma_0$ and hardening modulus
$E_\mathrm{h}$.
These conditions express that the von Mises stress should never exceed the yield stress and that plastic deformation is irreversible.
We assume an associated flow rule, so the plastic strain increment is expressed as:
\begin{align}
  \Delta\tens\epsilon^\mathrm{p} =
  \frac{3\Delta{e}^\mathrm{p}}{2f_\mathrm{y}(\tens\sigma + \calC:\Delta\tens \epsilon)}\tens s(\tens\sigma + \calC:\Delta\tens \epsilon).
\end{align}
The usual return-mapping algorithm~\autocite{simoComputationalInelasticity1998}
is used to compute $\Delta e^\mathrm{p}$.
Solving the coupled elastic-plastic contact problem is done with a fixed-point
iterative
approach~\autocite{jacqDevelopmentThreeDimensionalSemiAnalytical2002,frerotFourieracceleratedVolumeIntegral2019}
accelerated with an Anderson mixing
procedure~\autocite{andersonIterativeProceduresNonlinear1965,eyertComparativeStudyMethods1996}.
The code, available as a supplementary material, is built on top of the
open-source high-performance contact library \textsc{Tamaas}~\autocite{frerotTamaasLibraryElasticplastic2020}.

For reference, we also used the surface plasticity approach of
\textcite{almqvistDryElastoplasticContact2007} \changed{and a rigid-plastic,
  as known as a bearing area model}. \changed{The former} solves an elastic contact
problem with the constraint that the normal surface pressure be less than or
equal to the indentation hardness
$p_\mathrm{m}$~\autocite{taborHardnessMetals1951}, correcting the counterface so
that the constraint is satisfied. We use this correction as the residual plastic
displacement~\autocite{almqvistDryElastoplasticContact2007,weberMolecularProbesReveal2018}.
The choice of the hardness $p_\mathrm{m}$ is a debated question: While
$p_\mathrm{m} \approx 3\sigma_\mathrm{y}$ is a common choice stemming from
spherical indentation~\autocite{taborHardnessMetals1951}, finite element
simulations of
sinusoidal~\autocite{gaoBehaviorElasticPerfectly2006,ghaedniaReviewElasticPlastic2017}
and rough surfaces~\autocite{peiFiniteElementModeling2005} report values in the
order of $6\sigma_\mathrm{y}$, and a quantitative agreement with the $J_2$ model
requires a surface-specific adjustment of
$p_m$~\autocite{frerotCrackNucleationAdhesive2020}. Here we \changed{nonetheless} set
$p_\mathrm{m} = \changed{3}\sigma_\mathrm{y}$. The model is hereafter referred to as the saturated
plasticity model.
\changed{Finally, the rigid-plastic, or bearing area approach, consists in solving the contact entirely without elastic deformation, by finding a plane intersecting the rough counterface such that the intersection area equals the normal force, $F_\mathrm{n}$, divided by $p_\mathrm{m}$.}
Note that \changed{these two models corresponds roughly} to historic plastic
contact models by~\textcite{holmElectricContactsTheory2000}.

\subsection{Roughness Statistics}
Gaussian, self-affine rough surfaces \changed{with a period $L$}, are generated using a
random-phase algorithm~\autocite{wuSimulationRoughSurfaces2000,ramisettiAutocorrelationFunctionIsland2011,perssonNatureSurfaceRoughness2005,jacobsQuantitativeCharacterizationSurface2017} for each indentation step.
They follow the \changed{isotropic} power-spectral density (PSD)
\begin{align}
  \left|\mathcal{F}[h_k]\right|^2
  & =
  \begin{cases}
    C & \quad \text{if } q_0 \leq q \leq q_1,\\
    C\left(\frac{q}{q_1}\right)^{-2(H+1)} & \quad \text{if } q_1 \leq q \leq q_2,\\
    0 & \quad \text{otherwise,}
  \end{cases}
\end{align}
where $\mathcal{F}$ is the Fourier transform, $q_i$ are angular wavenumbers for
the cutoffs in the surface spectrum, $H$ is the Hurst exponent, and $C$ an
unspecified constant, adjusted such that the root-mean-square of surface slopes
has the desired expected value.

The residual displacement resulting from plastic deformation at step $k$ is
noted $\tens u^\mathrm{p}_k$ and is computed using a Green's function~\autocite{buiRemarksFormulationThreedimensional1978}, \changed{which accounts for ``spring-back'' effects,}
\begin{align}
  \tens u^\mathrm{p}_k(x) = \int_\body{\tens \nabla_y \tens U(x,
  y):\calC:\tens\epsilon^\mathrm{p}_k\, \mathrm{d}V_y},
\end{align}
where $\tens U$ is the Mindlin
tensor~\autocite{mindlinForcePointInterior1936,frerotMindlinFundamentalSolution2018,frerotFourieracceleratedVolumeIntegral2019}.
We note $h^\mathrm{p}_k$ the trace on $\partial \body$, the boundary of $\body$, of the component of
$\tens u^\mathrm{p}_k$ normal to the surface: It is the emerging surface
roughness due to successive indentations.

Be it from plastic indentation or from synthetic generation, we derive from a
rough surface $h$ the following statistical
properties~\autocite{jacobsQuantitativeCharacterizationSurface2017},
\begin{align}
  \left(h^{(\alpha)}_\mathrm{rms}\right)^2 & = \frac{1}{\changed{L^2}}\int_{\changed{A_L}}{|\nabla^\alpha h|^2\,\mathrm{d}A},\\
  h_\mathrm{rms} & = h^{(0)}_\mathrm{rms},\quad
  h'_\mathrm{rms} = h^{(1)}_\mathrm{rms},\quad
  h''_\mathrm{rms} = \frac{1}{2} h^{(2)}_\mathrm{rms},\\
  s & = \frac{1}{h_\mathrm{rms}^3\changed{L^2}}\int_{\changed{A_L}}{(h - \langle h
      \rangle)^3\,\mathrm{d}A},
\end{align}
where \changed{$A_L = [0, L]^2$}, $\nabla^\alpha$ is the fractional Riesz derivative, computed in the Fourier domain, which links $h^{(\alpha)}_\mathrm{rms}$ to Nayak's definition of PSD moments~\autocite{nayakRandomProcessModel1971}. The parameters $h_\mathrm{rms}$, $h'_\mathrm{rms}$ and $h''_\mathrm{rms}$ are the root-mean-square (RMS) of heights, RMS of slopes and RMS of curvatures, respectively. The parameter $s$ is the \changed{skewness of the surface height distribution}.

Rather than reporting the curvature directly, we report it in form of Nayak's parameter~\autocite{nayakRandomProcessModel1971},
\begin{equation}
  \alpha = 4(h_\mathrm{rms} h_\mathrm{rms}^{\prime\prime} / h_\mathrm{rms}^{\prime 2})^2.
\end{equation}
The advantage of using Nayak's parameter is that it removes the influence of changes in roughness amplitude.

Since rough surfaces are multi-scale objects, we use a multi-scale quantitative
roughness analysis in addition to the scalar measures defined above.
As one of the most common statistical descriptors, we report the \changed{isotropic} PSD $\phi(q) = |\mathcal F[h]|^2$.
We also apply a variable bandwidth method (VBM) in form of
detrended fluctuation
analysis~\autocite{cannonEvaluatingScaledWindowed1997,pengMosaicOrganizationDNA1994,pengQuantificationScalingExponents1995,jacobsQuantitativeCharacterizationSurface2017}.
Specifically, 
the surface with side length $L$ is divided into square patches of side length $\ell$
(see inset in~\cref{fig:scale_dependent} for an illustration). Each patch is
then independently detrended by subtracting the plane that minimizes $h_\mathrm{rms}$ in that patch. Evaluating the
above roughness parameters on the detrended patches for decreasing $\ell$ gives
length-scale dependent metrics.
Both PSD and VBM allow the study of a surface's potential
self-affinity, with self-affine surfaces following $\phi(q)\propto q^{-2-2H}$ or $h_\mathrm{rms}(\ell)\propto\ell^H$ over the self-affine bandwidth.

We also compute the dissipated plastic energy $d^\mathrm{p}_k$ during successive
indentations, given by
\begin{equation}
d^\mathrm{p}_k = \sum_{i=1}^k\int_{\changed{\body_L}}{\left(\tens\sigma_i:\Delta{\tens\epsilon}^\mathrm{p}_i -
                E_\mathrm{h}e^\mathrm{p}_i\Delta e^\mathrm{p}_i\right)\mathrm{d}V},
\end{equation}
for the $J_2$ model \changed{($\body_L = A_L\times [0, +\infty[$)}, and the irrecoverable energy, which includes the dissipated plastic energy and the stored elastic energy due to residual stresses, and can be computed directly on the surface for both models:
\begin{equation}
\delta^\mathrm{p}_k = \sum_{i=1}^k\int_{\changed{A_L}}{p_i\Delta{h}^\mathrm{p}_i\mathrm{d}A}.\\
\end{equation}
Note that the saturated model lacks a description of eigenstresses~\autocite{frerotCrackNucleationAdhesive2020}, thus $\delta^\mathrm p_k\equiv d^\mathrm p_k$ is the dissipated plastic energy for this model.
 
\section{Results}
We simulate the repeated indentation of 90 \changed{different} surfaces with the same statistical properties and the following PSD parameters: $q_0L/2\pi = 2$, $q_1L/2\pi = 2$, $q_2L/2\pi = 128$, $H \in \{0.5, 0.8\}$, $h'_\mathrm{rms} = 0.4$, where $L$ is the horizontal period of the surfaces. \changed{Since the computational cost of the saturated model is significantly lower than the $J_2$ model, we simulate it in 270 repeated indentation steps.}
Contact is made against a half-space with material properties $\nu = \changed{0.3}$, $\sigma_0 = 0.1 E$, $E_h = 0.1 E$, with $E$ the Young modulus. All properties are nondimensionalized by $E$ and by the horizontal period $L$ and only those parameters control the outcome of the simulations.

\begin{figure*}
  \centering
  \includegraphics{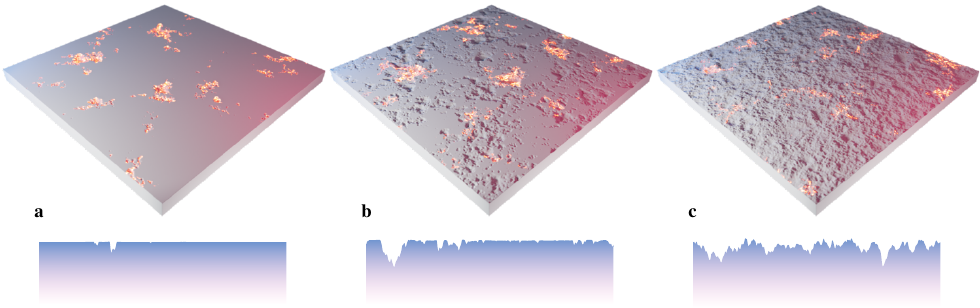}
  \caption{Evolution of surface roughness through repeated elastic-plastic
    indentation. Highlighted areas on the three-dimensional views show where plastic energy is
    dissipated at the specific step that is shown. The panels show the topography after (a) $k = 1$, (b) $10$ and (c)
    $90$ repeated contacts. The bottom row shows the evolution of one-dimensional cross-sectional
    profiles of the above surfaces. After sufficient number of iterations, no undeformed (flat) areas can be
    identified in the surface.}\label{fig:concept}
\end{figure*}

\Cref{fig:concept} shows a rendered view of the emerging surface roughness due to the accumulation of plastic residual deformation, for a normal force of $F_\mathrm{n}/(E h'_\mathrm{rms}\changed{L^2}) = 0.025$ \changed{(which corresponds to a contact area ratio of 7\%)}, and a discretization size of  729$\times$729$\times$55 points for a domain size of $L\times L\times 0.075L$ (total of 175M degrees-of-freedom).
The highlighted areas in~\cref{fig:concept} show where plastic energy is dissipated on the surface for the current indentation step.
We also show select one-dimensional cross-sectional profiles of these topography maps below them.
After enough steps, no remaining ``undeformed'' area of the surface can be identified by eye, and the surface appears entirely rough from plasticity.

\begin{figure*}
  \includegraphics{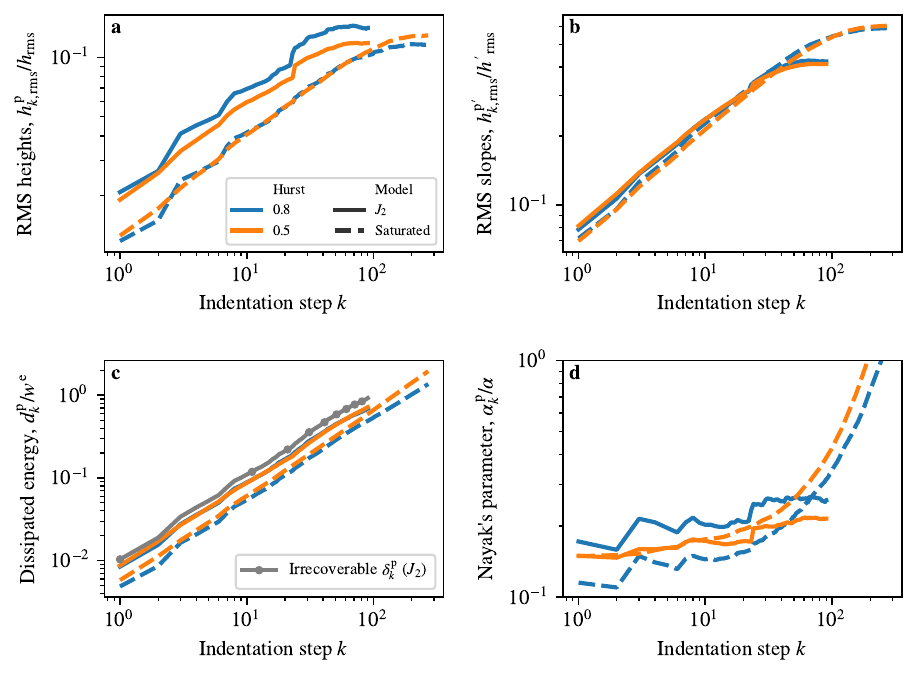}
  \caption{Evolution of the normalized root-mean-square (RMS) of (a) heights, (b) slopes, (c) dissipated and irrecoverable
    energy, and (d) Nayak's parameter with indentation
    step. While, \changed{for the $J_2$ model,} the RMS of heights and slopes show similar evolution, with a steady-state after $\sim 40$ contact cycles, Nayak's parameter converges earlier and the dissipated energy does not reach a steady-state value within $90$ contacts. \changed{The scaling of the saturated results with $k$ is similar to the $J_2$ model, with a steady state reached after a larger number of iterations}. Nayak's parameter (d) differs qualitatively between calculations with $J_2$ and saturated plasticity models due to slope discontinuities \changed{in the residual displacement}.}\label{fig:rms}
\end{figure*}

We first evaluate macroscopic scalar roughness parameters on the emerging roughness.
\Cref{fig:rms} shows the evolution, for the indented roughness, of
the root-mean-square of heights $h^\mathrm{p}_{k,\mathrm{rms}}$ (panel a), the RMS of
slopes $h^{\mathrm{p}\prime}_{k,\mathrm{rms}}$ (b), the dissipated energy
$d^\mathrm p_k$ and the irrecoverable energy $\delta^\mathrm{p}_k$ normalized by the elastic conforming energy of the counterface
$w^e = [h^{(1/2)}_\mathrm{rms}]^2 E/(4(1-\nu^2)) $ (c), and Nayak's parameter, $\alpha^\mathrm p$ (d).
The dynamics of the first three parameters is similar in the two plasticity models, $J_2$ (von Mises), and saturated.
Both RMS metrics initially scale as a power-law with an exponent close to $1/2$ before plateauing at steady-state values, with the $J_2$ model transitioning earlier to lower values than the saturated model.
\changed{Interestingly, the saturated model, which does not have hardening, also reaches a plateau.}
\changed{The overall scaling of RMS metrics is independent of $p_\mathrm{m}$ (not shown), and $p_\mathrm{m}$ only controls the contact area at each indentation: A lower contact ratio delays the appearance of a steady-state roughness}.
While energy dissipation in the saturated models appears to be constant per indentation step, the $J_2$ model shows lower energy dissipation rates after the first $\sim 20$ contact cycles.
The irrecoverable energy shows a trend similar to the dissipated energy.
It has a larger value, which indicates a build-up of residual stresses with successive indentations.
Residual stresses require work for their formation, and while the stored elastic energy is in principle reversible, it is not recoverable in practice.
The ratio $\delta^\mathrm{p}_k / d^\mathrm{p}_k$ increases from 1.22 to 1.34 with $k$, reflecting an increase in the proportion of elastic stored energy in $\delta^\mathrm{p}_k$, consistent with hardening in the $J_2$ model.

As a geometric measure, Nayak's parameter evolves differently in the $J_2$ and the saturated plasticity model.
In the saturated plasticity model, residual displacements are non-zero only in areas where the contact pressure $p = p_\mathrm{m}$.
This criterion creates a slope discontinuity at the edge of zones where $p = p_\mathrm{m}$ which dominates the high-frequency part of the PSD.
They are reflected in large values of $h''_\mathrm{rms}$, contributing to the sustained increase in $\alpha_k^\mathrm p$ in the saturated model.

The influence of the Hurst exponent $H$ disappears when we nondimensionalize lengths, slopes and energies by RMS heights, slopes and conforming elastic energy, respectively, of the counterface.
However, normalization does not collapse the steady-state value of RMS heights and Nayak's parameter, indicating that details of the counterface self-affine structure non-trivially affect the long-wavelength part of the emerging roughness spectrum.

\begin{figure*}
  \centering
  \includegraphics{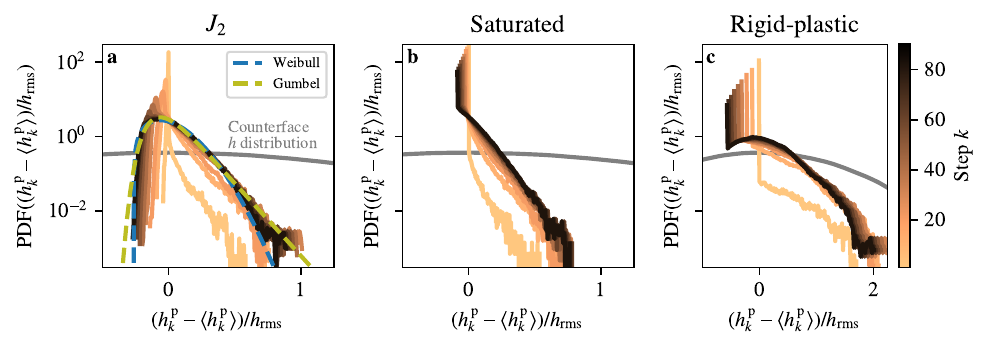}
  \caption{Evolution of the probability density function (PDF) of residual roughness
    heights for (a) the $J_2$, (b) the saturated plasticity model \changed{and (c) the rigid-plastic model}. The peak corresponds to the residual deformation in untouched areas. As more areas come in contact, the peak broadens and eventually disappears. The height distribution then tends towards a non-Gaussian, asymmetric form, which resembles common extreme value distributions.}\label{fig:pdf}
\end{figure*}

Having established the evolution of macroscopic parameters with contact cycle, we now detail the
evolution of the distribution of heights.
\Cref{fig:pdf} shows, for the $J_2$
(a), saturated (b), \changed{and rigid-plastic (c) models}, the height probability density function of the
emerging roughness through the indentation steps, for the counterface with
$H = 0.8$.

While the distributions are always non-Gaussian, there is a clear
evolution in the shape of the distribution with $k$: For low indentation count,
the distribution obtained from the $J_2$ model is dominated by a well-defined peak, similar to a rectified
distribution, with an exponential tail. This peak occurs for a height which
corresponds to the residual displacement in the zones that have not been in
contact yet. The peak has a finite width due to the non-local influence of
subsurface residual strains.
As more surfaces come in contact with increasing
$k$, the ``undeformed'' area of the surface shrinks: the peak height decreases
and its breadth increases due to a larger variation of residual displacements,
caused by ever-increasing, in number and size, plastic zones. The peak
eventually disappears, and the height distribution settles on a more regular
shape.

Unlike the $J_2$ model, the saturated model gives a true rectified distribution, where the peak in \cref{fig:pdf}a becomes a Dirac distribution in \cref{fig:pdf}b, whose magnitude is given by the remaining (truly) undeformed area. The tail of the distribution at large (positive) heights is similar for both models, \changed{although the saturated model has not reached a steady-state RMS of heights at the last iteration shown in \cref{fig:pdf}b}. \changed{Similarly to the saturated, the rigid-plastic model (\cref{fig:pdf}c) also shows a rectified distribution, except that the residual height range is larger by a factor of approximately 2. This is due to the initial elastic deformation needed to overcome the plastic yield stress, absent in the rigid-plastic approach.}
%

\Cref{fig:pdf} shows fits of the simulation data with two extreme value distributions: the
Weibull\footnote{cumulative distribution
  $\mathrm{cdf}(x) = 1 - \exp((-x / \lambda)^\kappa)$, \changed{$\lambda = 0.3$, $\kappa = 1.9$}} distribution, which
\textcite{silvasabinoEvolutionRealContact2023} used in an elastic contact
analysis of non-Gaussian surfaces, and the Gumbel
distribution\footnote{cumulative distribution
  $\mathrm{cdf}(x) = \exp(-\exp(-x/\beta))$, \changed{$\beta = 1/9$}}. We use extreme value distributions because, \changed{owing to the relatively low contact area and to long-range elastic interactions---which makes contact near large asperities unlikely}, only the largest asperities
of the counterface come into contact, leading to a biased sampling of the
counterface roughness by the deformable elastic-plastic manifold, \changed{which retains deformation from the roughness it ``saw'' in contact}.
Neither of the two distribution functions appears to be a perfect fit, but the Gumbel distribution captures the largest values better than Weibull. This clearly shows that the overall distribution has non-Gaussian tails.
\changed{We note that the saturated plasticity model also has non-Gaussian tails, as it samples the peaks of the counterface. In contrast, the rigid-plastic (bearing area) model simply reflects the Gaussian distribution of the counterface.}

\begin{figure}
  \centering
  \includegraphics{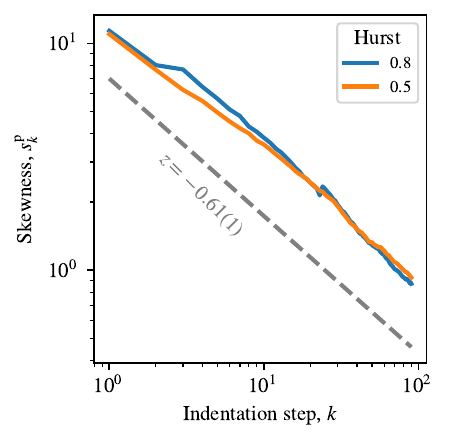}
  \caption{Evolution of the skewness of the height distribution with indentation step $k$ for the
    $J_2$ plasticity model. The skewness follows a power-law $k^{z}$, whose dynamics exponent $z$ is
    independent of the counterface's Hurst exponent. This exponent describes the dynamics of the roughness evolution.}\label{fig:skewness}
\end{figure}

From the prior results is it clear that the height distribution is not symmetric, which we now characterize from its skewness.
\Cref{fig:skewness} shows, for the $J_2$ model only, that the skewness of the height distribution, $s^\mathrm p_k$, decreases with $k$ as a power-law that is independent of the counterface's Hurst exponent.
Unlike the RMS of heights or slopes, there is no observable transition from a power-law to a constant regime.
We obtain a power-law exponent of \changed{$z = -0.61(1)$} from a least-squares fit to the logarithm of the data.
Although $-1 < z < 0$ indicates a slow evolution towards a symmetric height distribution, the skewness may converge to a finite value if the surface hardens enough to prevent further plastic dissipation, which would indicate elastic shakedown.
Nonetheless, as we show next, the exponent $z$ is representative of the evolution dynamics of the roughness at all scales.

\begin{figure*}
  \centering
  \includegraphics{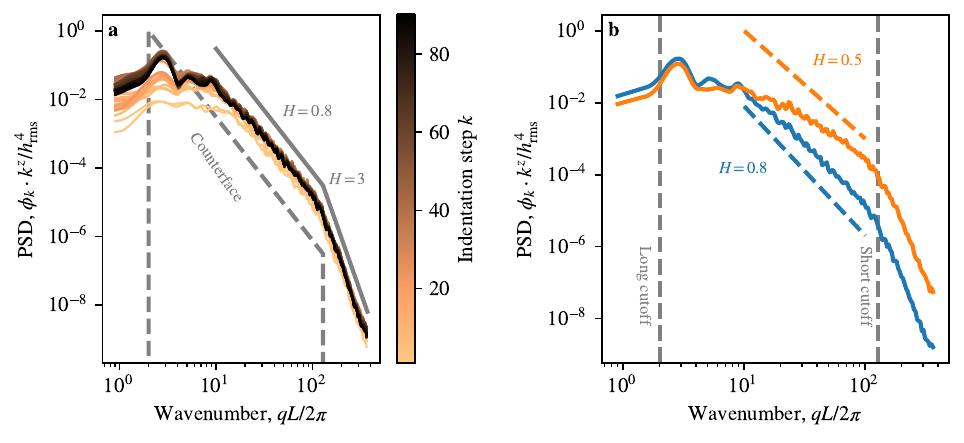}
  \caption{(a) Evolution of the power-spectral density (PSD) with indentation step for $H = 0.8$, and (b) final power-spectral density for both $H = 0.5$ and $H = 0.8$. Scaling the PSD with $k^{z}$ collapses all curves to a single master curve. The master curve is constant at low wavenumbers (large wavelengths) and has two distinct power-law regimes at intermediate and small wavelengths. The intermediate scaling regime follows self-affine scaling of the counterface only for $H=0.8$. The topographies obtained after indentation with $H=0.5$ and $H=0.8$ show universal $q^{-8}$ scaling beyond the counterface's short wavelength cutoff.}\label{fig:psd}
\end{figure*}
\Cref{fig:psd} (a) shows, for each indentation step $k$ and for $H = 0.8$,
the power-spectral density (PSD) of the emerging roughness scaled by $k^z$:
individual PSDs then collapse onto a singular master curve above a wavenumber of $\sim$ 10.
The scaled PSD
shows a constant regime up to that wavenumber, where it transitions to a
regime with a Hurst exponent of $H^\mathrm p \approx 0.8$, which corresponds to
the counterface. Beyond the counterface's short wavelength cutoff (at wavenumber
128), the PSD scales as $q^{-8}$, which corresponds to $H^\mathrm p = 3$. This
scaling regime is due to the mechanical response at small scales: \cref{fig:psd}b
shows the PSDs of the last indentation for counterfaces with $H = 0.5$
and $H = 0.8$, and both have identical scaling beyond the short wavelength
cutoff. The self-affine regime 
does not exist in the simulations we conducted in the case of a counterface with $H = 0.5$. The transition from a
constant (white noise) to power-law (self-affine or elastic response) occurs on a range of scales too large to be captured within
the counterface's spectral bandwidth ($q_2 / q_1 = 64$). We discuss
below whether both resulting surfaces are in fact self-affine in the regime
where the counterface is self-affine.

Note that only ``vertical'' scaling of the PSD was required to obtain a
collapse, indicating that the final surface's scaling behavior is already
encoded in the first indentations. This is corroborated by the behavior of
Nayak's parameter in \cref{fig:rms} which is independent of the PSD's absolute
magnitude, and shows little variation with $k$.

\begin{figure*}
  \centering
  \includegraphics{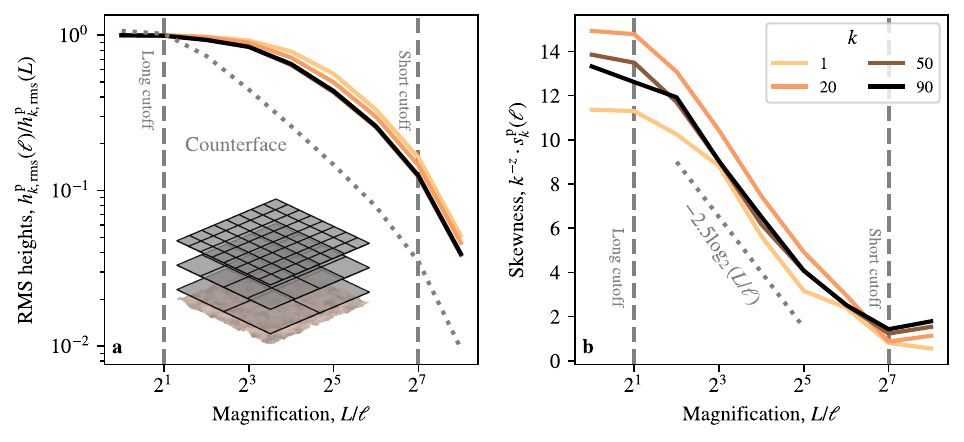}
  \caption{Variable bandwidth (VBM) analysis of (a) RMS heights and (b) skewness, for the simulations with a counterface with $H = 0.8$. The inset in the left plot illustrates the successive surface subdivision which the VBM employs.}\label{fig:scale_dependent}
\end{figure*}

In \cref{fig:scale_dependent}, we show a multi-scale VBM analysis---illustrated by the schematic inset in (a)---of two roughness parameters: the scale-dependent RMS of heights (a), $h^\mathrm p_{k,\mathrm{rms}}(\ell)$, and skewness (b), $s^\mathrm p_k(\ell)$, where $\ell$ is the window size.
An infinite self-affine surface shows a power-law behavior of $h_\mathrm{rms}(\ell)\propto \ell^H$.
Due to the restricted bandwidth of the spectrum, the counterface follows this scaling approximately (but not perfectly).
In contrast to this,  the emerging roughness is clearly not self-affine.

The skewnews, shown in \cref{fig:scale_dependent}b, is normalized by $k^{-z}$.
The collapse of skewness for all scales of curves with different $k$ corroborates the observation that $z$ describes the roughness evolution dynamics at all scales. As magnification increases, the skewness decreases, i.e. the height distributions become more symmetric at smaller scales.

\begin{figure}
  \centering
  \includegraphics{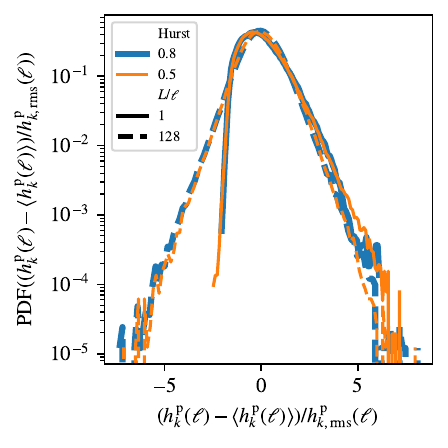}
  \caption{Height distributions after $90$ repeated contacts for magnifications $L / \ell = 1$ and
    $L / \ell = 128$. The plot shows the results obtained for counterfaces with $H = 0.5$ and $H = 0.8$. Normalization by
    $h^\mathrm p_\mathrm{rms}(\ell)$ shows that the Hurst exponent does not
    affect the distribution shape. At small scales, the height distribution
    appears symmetric but non-Gaussian with exponential tails.}\label{fig:small_scale}
\end{figure}

This behavior is directly shown in~\cref{fig:small_scale}, which depicts the height distribution at magnifications $L/\ell=1$ and $128$ for both values of the Hurst exponent at the last step $k = 90$.
The plot shows that the Hurst exponent has no influence on the shape of the height distribution.
At large magnification, corresponding to a small wavelength cutoff of the counterface, the height distribution is symmetric, although non-Gaussian, with approximately exponential tails.
This is roughly consistent with the Gumbel tails observed at the length scale $L$ of the overall simulations.


\section{Discussion}

Our simulation show that some roughness properties, such as the RMS height or slope, reach a steady-state long before the surface actually ``shakes downs'', in the sense that it stops deforming plastically and reacts only elastically and reversibly.
Indeed, our simulations never reach a true shakedown, but our analysis of the dissipated energy indicates that the rate of plastic deformation decreases with contact cycle, but only for the $J_2$ model that explicitly considers hardening.
Shakedown is never reached in the saturated plasticity model without hardening, even though the roughness may appear to have reached a state-state.

In~\cref{fig:rms}a, the RMS of heights for both models initially scales as $\sqrt{k}$. Despite memory (from plasticity), and long-range correlations (from elasticity), the RMS of heights behaves similarly to a simple ballistic deposition model at short times~\cite{barabasiFractalConceptsSurface1995}, indicating that early iterations are essentially uncorrelated.
This can also be observed in the dissipated energy, as early indentations are unaffected by hardening, while later indentations occur on already hardened areas.
\changed{However, since a perfect-plasticity hypothesis also yields a steady-state, we conclude that hardening only plays a minor role in the number of iterations required to observe a steady-state.}

\changed{Another} interesting aspect of our simulations is the emergence of non-Gaussian height distributions.
Non-Gaussianity is manifested in skewness as well as exponential tails.
The skewness has a well-defined monotonous dynamics described by a simple power-law $s\propto k^{z}$ with $-1<z<0$.
This type of dynamics implies that the emerging rough geometry \changed{slowly} becomes symmetric.

More insights are obtained by looking at the skewness as a function of scales.
\Cref{fig:scale_dependent}b clearly shows that the small-scale roughness is more
symmetric than the larger scales, and it appears that a factor of $k^{-z}$
collapses the dynamic evolution of the skewness even at small scales.
This seems to indicate that during repeated indentation, the small-scale roughness diffuses towards larger scales, or in other words that the small-scales are representative of height-distribution at larger scales obtained at later indentation cycles.
The height distributions hence become symmetric in the asymptotic steady-state.

However, the small-scale roughness does not follow a Gaussian distribution but has exponential tails (see \cref{fig:small_scale}).
The above argument seems to indicate, that even in the steady-state where height distributions are symmetric they are still not Gaussian.
It is noteworthy that height distributions with exponential tails are typically assumed in independent-asperity models of rough contacts to make them analytically tractable~\cite{greenwoodContactNominallyFlat1966,perssonSlidingFriction2000}.
The height distribution has also been shown to have implications in the way the true contact area depends on
the applied normal force~\autocite{silvasabinoEvolutionRealContact2023}, which
could have consequences in contact
sealing~\autocite{dappSelfAffineElasticContacts2012,shvartsTrappedFluidContact2018}
and wear particle
formation~\autocite{frerotMechanisticUnderstandingWear2018,popovAdhesiveWearParticle2018,brinkParameterfreeMechanisticModel2021}, although some of these applications would also see the development of anisotropic roughness, due to sliding for example~\autocite{candelaMinimumScaleGrooving2016}.
%
%


\section{Summary \& Conclusions}

In summary, we have shown how surface roughness of an initially flat elastic-plastic substrate evolves during repeated contact with a rigid self-affine counterface. The resulting rough topography is neither self-affine nor Gaussian, even if the rigid body has those properties.
Common scalar roughness measures, such as the RMS height, converge quickly to a steady-state value. Yet, subsurface plastic deformation still occurs long after the topography appears to have shaken-down geometrically. The topography continues to evolve, but this can only be seen at intermediate scales obtained from a scale-dependent geometric analyses, such as PSD.

The question of elastic shakedown has preoccupied tribologists for more than half a century~\autocite{greenwoodContactNominallyFlat1966,kapoorSteadyStateSliding1994,vakisModelingSimulationTribology2018}.
Shakedown occurs during complex phenomena such friction~\autocite{flicekDeterminationShakedownLimit2015}, fretting~\autocite{fouvryElasticPlasticShakedown2001} and rolling~\autocite{berthePlasticDeformationRough2014}.
Our simplified simulations are directly representative of processes such as sheet metal forming, where a die is brought into repeated contact with a counterface~\autocite{loSurfaceRougheningContact1999}.
However, they certainly also hold insights for roughness evolution during sliding but clearly cannot explain the development of anisotropic topographies~\autocite{candelaMinimumScaleGrooving2016}.
Our results contribute to the recent discussion on the emergence of surface roughness on solid but deformable bodies~\cite{milaneseEmergenceSelfaffineSurfaces2019,hinkleEmergenceSmallscaleSelfaffine2020,aghababaeiHowRoughnessEmerges2022}.

\paragraph{Acknowledgments}
The authors acknowledge the financial support of the Deutsche
Forschungsgemeinschaft (DFG Grant No. 461911253, AWEARNESS). Computational
resources from NEMO, at the University of Freiburg (DFG Grant No. INST 39/963-1
FUGG) are also acknowledged. The library
\textsc{Tamaas}~\autocite{frerotTamaasLibraryElasticplastic2020} was used for
contact simulation, roughness post-processing was done with \textsc{Tamaas} and
\textsc{SurfaceTopography}~\autocite{rottgerContactEngineeringCreate2022}.
\textsc{Blender} and
\textsc{Matplotlib}~\autocite{hunterMatplotlib2DGraphics2007} were used for
visualization.

\paragraph{Author Contributions}
L.F. and L.P. designed the study, discussed results and edited the final manuscript. L.F. carried out calculations, analyzed the data and wrote the first manuscript draft.

\paragraph{Data Availability}
Simulation and figure codes are available on Zenodo~\autocite{frerotSupplementaryCodesData2023}.

\paragraph{Competing Interests} The authors declare no competing interests.


\printbibliography

@article{aghababaeiHowRoughnessEmerges2022,
  title = {How Roughness Emerges on Natural and Engineered Surfaces},
  author = {Aghababaei, Ramin and Brodsky, Emily E. and Molinari, Jean-François and Chandrasekar, Srinivasan},
  date = {2022-12-01},
  journaltitle = {MRS Bulletin},
  shortjournal = {MRS Bulletin},
  volume = {47},
  number = {12},
  pages = {1229--1236},
  issn = {1938-1425},
  doi = {10.1557/s43577-022-00469-1},
  urldate = {2023-07-21},
  langid = {english}
}

@article{almqvistDryElastoplasticContact2007,
  title = {On the Dry Elasto-Plastic Contact of Nominally Flat Surfaces},
  author = {Almqvist, A. and Sahlin, F. and Larsson, R. and Glavatskih, S.},
  date = {2007-04},
  journaltitle = {Tribology International},
  volume = {40},
  number = {4},
  pages = {574--579},
  issn = {0301679X},
  doi = {10.1016/j.triboint.2005.11.008},
  urldate = {2018-01-31},
  langid = {english}
}

@article{andersonIterativeProceduresNonlinear1965,
  title = {Iterative {{Procedures}} for {{Nonlinear Integral Equations}}},
  author = {Anderson, Donald G.},
  date = {1965-10},
  journaltitle = {Journal of the ACM},
  shortjournal = {J. ACM},
  volume = {12},
  number = {4},
  pages = {547--560},
  issn = {0004-5411, 1557-735X},
  doi = {10.1145/321296.321305},
  urldate = {2023-05-30},
  langid = {english}
}

@book{barabasiFractalConceptsSurface1995,
  title = {Fractal {{Concepts}} in {{Surface Growth}}},
  author = {Barabási, A.- L. and Stanley, H. E.},
  date = {1995-04-13},
  edition = {1},
  publisher = {{Cambridge University Press}},
  doi = {10.1017/CBO9780511599798},
  urldate = {2023-08-22},
  isbn = {978-0-521-48308-7 978-0-521-48318-6 978-0-511-59979-8}
}

@article{berthePlasticDeformationRough2014,
  title = {Plastic Deformation of Rough Rolling Contact: {{An}} Experimental and Numerical Investigation},
  shorttitle = {Plastic Deformation of Rough Rolling Contact},
  author = {Berthe, L. and Sainsot, P. and Lubrecht, A. A. and Baietto, M. C.},
  date = {2014-04-15},
  journaltitle = {Wear},
  shortjournal = {Wear},
  volume = {312},
  number = {1–2},
  pages = {51--57},
  issn = {0043-1648},
  doi = {10.1016/j.wear.2014.01.017},
  urldate = {2017-05-01}
}

@article{binderWiderstandKontakten1912,
  title = {Der Widerstand von Kontakten},
  author = {Binder, Ludwig},
  date = {1912},
  journaltitle = {Elektrotechnik und Maschinenbau},
  volume = {30},
  pages = {781--788}
}

@article{bonnetImplicitBEMFormulations1996,
  title = {Implicit {{BEM}} Formulations for Usual and Sensitivity Problems in Elasto-Plasticity Using the Consistent Tangent Operator Concept},
  author = {Bonnet, Marc and Mukherjee, Subrata},
  date = {1996-12},
  journaltitle = {International Journal of Solids and Structures},
  shortjournal = {International Journal of Solids and Structures},
  volume = {33},
  number = {30},
  pages = {4461--4480},
  issn = {0020-7683},
  doi = {10.1016/0020-7683(95)00279-0},
  urldate = {2017-01-19}
}

@article{brinkParameterfreeMechanisticModel2021,
  ids = {brinkParameterfreeMechanisticModel2020},
  title = {A Parameter-Free Mechanistic Model of the Adhesive Wear Process of Rough Surfaces in Sliding Contact},
  author = {Brink, Tobias and Frérot, Lucas and Molinari, Jean-François},
  date = {2021-02-01},
  journaltitle = {Journal of the Mechanics and Physics of Solids},
  shortjournal = {Journal of the Mechanics and Physics of Solids},
  volume = {147},
  eprint = {2004.00559},
  eprinttype = {arxiv},
  pages = {104238},
  issn = {0022-5096},
  doi = {10.1016/j.jmps.2020.104238},
  urldate = {2021-01-19},
  langid = {english}
}

@article{buiRemarksFormulationThreedimensional1978,
  title = {Some Remarks about the Formulation of Three-Dimensional Thermoelastoplastic Problems by Integral Equations},
  author = {Bui, H. D.},
  date = {1978-01-01},
  journaltitle = {International Journal of Solids and Structures},
  shortjournal = {International Journal of Solids and Structures},
  volume = {14},
  number = {11},
  pages = {935--939},
  issn = {0020-7683},
  doi = {10.1016/0020-7683(78)90069-0},
  urldate = {2018-02-06}
}

@article{campanaElasticContactSelfaffine2008,
  title = {Elastic Contact between Self-Affine Surfaces: Comparison of Numerical Stress and Contact Correlation Functions with Analytic Predictions},
  shorttitle = {Elastic Contact between Self-Affine Surfaces},
  author = {Campañá, Carlos and Müser, Martin H. and Robbins, Mark O.},
  date = {2008},
  journaltitle = {Journal of Physics: Condensed Matter},
  shortjournal = {J. Phys.: Condens. Matter},
  volume = {20},
  number = {35},
  pages = {354013},
  issn = {0953-8984},
  doi = {10.1088/0953-8984/20/35/354013},
  urldate = {2016-07-18},
  langid = {english}
}

@article{candelaMinimumScaleGrooving2016,
  title = {The Minimum Scale of Grooving on Faults},
  author = {Candela, Thibault and Brodsky, Emily E.},
  date = {2016-08-01},
  journaltitle = {Geology},
  shortjournal = {Geology},
  volume = {44},
  number = {8},
  pages = {603--606},
  issn = {0091-7613, 1943-2682},
  doi = {10.1130/G37934.1},
  urldate = {2017-01-23},
  langid = {english}
}

@article{cannonEvaluatingScaledWindowed1997,
  title = {Evaluating Scaled Windowed Variance Methods for Estimating the {{Hurst}} Coefficient of Time Series},
  author = {Cannon, Michael J. and Percival, Donald B. and Caccia, David C. and Raymond, Gary M. and Bassingthwaighte, James B.},
  date = {1997-07-15},
  journaltitle = {Physica A: Statistical Mechanics and its Applications},
  shortjournal = {Physica A: Statistical Mechanics and its Applications},
  volume = {241},
  number = {3},
  pages = {606--626},
  issn = {0378-4371},
  doi = {10.1016/S0378-4371(97)00252-5},
  urldate = {2023-08-03},
  langid = {english}
}

@article{dappSelfAffineElasticContacts2012,
  title = {Self-{{Affine Elastic Contacts}}: {{Percolation}} and {{Leakage}}},
  shorttitle = {Self-{{Affine Elastic Contacts}}},
  author = {Dapp, Wolf B. and Lücke, Andreas and Persson, Bo N. J. and Müser, Martin H.},
  date = {2012-06-15},
  journaltitle = {Physical Review Letters},
  shortjournal = {Phys. Rev. Lett.},
  volume = {108},
  number = {24},
  pages = {244301},
  doi = {10.1103/PhysRevLett.108.244301},
  urldate = {2019-09-25}
}

@article{eyertComparativeStudyMethods1996,
  title = {A {{Comparative Study}} on {{Methods}} for {{Convergence Acceleration}} of {{Iterative Vector Sequences}}},
  author = {Eyert, V.},
  date = {1996-03-15},
  journaltitle = {Journal of Computational Physics},
  shortjournal = {Journal of Computational Physics},
  volume = {124},
  number = {2},
  pages = {271--285},
  issn = {0021-9991},
  doi = {10.1006/jcph.1996.0059},
  urldate = {2023-05-26},
  langid = {english}
}

@article{flicekDeterminationShakedownLimit2015,
  title = {Determination of the Shakedown Limit for Large, Discrete Frictional Systems},
  author = {Flicek, R. C. and Hills, D. A. and Barber, J. R. and Dini, D.},
  date = {2015-01-01},
  journaltitle = {European Journal of Mechanics - A/Solids},
  shortjournal = {European Journal of Mechanics - A/Solids},
  volume = {49},
  pages = {242--250},
  issn = {0997-7538},
  doi = {10.1016/j.euromechsol.2014.08.001},
  urldate = {2023-07-31},
  langid = {english}
}

@article{fouvryElasticPlasticShakedown2001,
  title = {An Elastic–Plastic Shakedown Analysis of Fretting Wear},
  author = {Fouvry, S. and family=Kapsa, given=Ph., given-i={{Ph}} and Vincent, L.},
  date = {2001-01-01},
  journaltitle = {Wear},
  shortjournal = {Wear},
  volume = {247},
  number = {1},
  pages = {41--54},
  issn = {0043-1648},
  doi = {10.1016/S0043-1648(00)00508-1},
  urldate = {2023-03-27},
  langid = {english}
}

@article{frerotCrackNucleationAdhesive2020,
  ids = {frerot_crack_2019},
  title = {Crack Nucleation in the Adhesive Wear of an Elastic-Plastic Half-Space},
  author = {Frérot, Lucas and Anciaux, Guillaume and Molinari, Jean-François},
  date = {2020-12-01},
  journaltitle = {Journal of the Mechanics and Physics of Solids},
  shortjournal = {Journal of the Mechanics and Physics of Solids},
  volume = {145},
  eprint = {1910.05163},
  eprinttype = {arxiv},
  pages = {104100},
  issn = {0022-5096},
  doi = {10.1016/j.jmps.2020.104100},
  urldate = {2020-09-04},
  langid = {english}
}

@article{frerotFourieracceleratedVolumeIntegral2019,
  title = {A {{Fourier-accelerated}} Volume Integral Method for Elastoplastic Contact},
  author = {Frérot, Lucas and Bonnet, Marc and Molinari, Jean-François and Anciaux, Guillaume},
  date = {2019-07-01},
  journaltitle = {Computer Methods in Applied Mechanics and Engineering},
  shortjournal = {Computer Methods in Applied Mechanics and Engineering},
  volume = {351},
  pages = {951--976},
  issn = {0045-7825},
  doi = {10.1016/j.cma.2019.04.006},
  urldate = {2019-05-06}
}

@article{frerotMechanisticUnderstandingWear2018,
  title = {A Mechanistic Understanding of the Wear Coefficient: {{From}} Single to Multiple Asperities Contact},
  shorttitle = {A Mechanistic Understanding of the Wear Coefficient},
  author = {Frérot, Lucas and Aghababaei, Ramin and Molinari, Jean-François},
  date = {2018-02-28},
  journaltitle = {Journal of the Mechanics and Physics of Solids},
  shortjournal = {Journal of the Mechanics and Physics of Solids},
  volume = {114},
  pages = {172--184},
  issn = {0022-5096},
  doi = {10.1016/j.jmps.2018.02.015},
  urldate = {2018-03-06}
}

@software{frerotMindlinFundamentalSolution2018,
  title = {The {{Mindlin Fundamental Solution}} - {{A Fourier Approach}}},
  author = {Frérot, Lucas},
  date = {2018-11-20},
  doi = {10.5281/zenodo.1492149},
  urldate = {2018-11-28},
  organization = {{Zenodo}}
}

@software{frerotSupplementaryCodesData2023,
  title = {Supplementary Codes and Data to "{{Elastic}} Shakedown and Roughness Evolution in Repeated Elastic-Plastic Contact"},
  author = {Frérot, Lucas and Pastewka, Lars},
  date = {2023-08-24},
  doi = {10.5281/zenodo.8280362},
  urldate = {2023-08-24},
  organization = {{Zenodo}}
}

@article{frerotTamaasLibraryElasticplastic2020,
  title = {Tamaas: A Library for Elastic-Plastic Contact of Periodic Rough Surfaces},
  shorttitle = {Tamaas},
  author = {Frérot, Lucas and Anciaux, Guillaume and Rey, Valentine and Pham-Ba, Son and Molinari, Jean-François},
  date = {2020-07-28},
  journaltitle = {Journal of Open Source Software},
  volume = {5},
  number = {51},
  pages = {2121},
  issn = {2475-9066},
  doi = {10.21105/joss.02121},
  urldate = {2020-07-29},
  langid = {english}
}

@article{gaoBehaviorElasticPerfectly2006,
  title = {The Behavior of an Elastic–Perfectly Plastic Sinusoidal Surface under Contact Loading},
  author = {Gao, Y. F. and Bower, A. F. and Kim, K. -S. and Lev, L. and Cheng, Y. T.},
  date = {2006-07-31},
  journaltitle = {Wear},
  shortjournal = {Wear},
  volume = {261},
  number = {2},
  pages = {145--154},
  issn = {0043-1648},
  doi = {10.1016/j.wear.2005.09.016},
  urldate = {2020-03-04},
  langid = {english}
}

@online{garcia-suarezRoughnessEvolutionInduced2023,
  title = {Roughness Evolution Induced by Third-Body Wear},
  author = {Garcia-Suarez, Joaquin and Brink, Tobias and Molinari, Jean-François},
  date = {2023-06-15},
  eprint = {2306.08993},
  eprinttype = {arxiv},
  eprintclass = {cond-mat},
  doi = {10.48550/arXiv.2306.08993},
  urldate = {2023-07-21},
  pubstate = {preprint}
}

@article{ghaedniaReviewElasticPlastic2017,
  title = {A {{Review}} of {{Elastic}}–{{Plastic Contact Mechanics}}},
  author = {Ghaednia, Hamid and Wang, Xianzhang and Saha, Swarna and Xu, Yang and Sharma, Aman and Jackson, Robert L.},
  date = {2017-11-14},
  journaltitle = {Applied Mechanics Reviews},
  shortjournal = {Appl. Mech. Rev},
  volume = {69},
  number = {6},
  pages = {060804-060804-30},
  issn = {0003-6900},
  doi = {10.1115/1.4038187},
  urldate = {2019-06-27}
}

@article{greenwoodContactNominallyFlat1966,
  title = {Contact of {{Nominally Flat Surfaces}}},
  author = {Greenwood, J. A. and Williamson, J. B. P.},
  date = {1966-12-06},
  journaltitle = {Proceedings of the Royal Society of London A: Mathematical, Physical and Engineering Sciences},
  volume = {295},
  number = {1442},
  pages = {300--319},
  issn = {1364-5021, 1471-2946},
  doi = {10.1098/rspa.1966.0242},
  urldate = {2016-02-18},
  langid = {english}
}

@article{hinkleEmergenceSmallscaleSelfaffine2020,
  title = {The Emergence of Small-Scale Self-Affine Surface Roughness from Deformation},
  author = {Hinkle, Adam R. and Nöhring, Wolfram G. and Leute, Richard and Junge, Till and Pastewka, Lars},
  date = {2020-02-14},
  journaltitle = {Science Advances},
  volume = {6},
  number = {7},
  pages = {eaax0847},
  publisher = {{American Association for the Advancement of Science}},
  doi = {10.1126/sciadv.aax0847},
  urldate = {2023-07-12}
}

@article{Holm1938-dh,
  title = {Eine Bestimmung der wirklichen Berührungsfläche eines Bürstenkontakes},
  author = {Holm, Ragnar},
  date = {1938},
  journaltitle = {Wissenschaftliche Veröffentlichungen aus den Siemens-Werken},
  volume = {17},
  number = {4},
  pages = {405--409},
  langid = {ngerman}
}

@article{Holm1938-va,
  title = {Über die auf die wirkliche Berührungsfläche bezogene Reibkraft},
  author = {Holm, Ragnar},
  date = {1938},
  journaltitle = {Wissenschaftliche Veröffentlichungen aus den Siemens-Werken},
  volume = {17},
  number = {4},
  pages = {400--404},
  langid = {ngerman}
}

@article{Holm1941-zc,
  title = {Beitrag zur Kenntnis der Reibung},
  author = {Holm, Ragnar},
  date = {1941},
  journaltitle = {Wissenschaftliche Veröffentlichungen aus den Siemens-Werken},
  volume = {20},
  number = {1},
  pages = {68--84},
  langid = {ngerman}
}

@book{holmElectricContactsTheory2000,
  title = {Electric Contacts: Theory and Applications},
  shorttitle = {Electric Contacts},
  author = {Holm, Ragnar},
  date = {2000},
  edition = {4th ed},
  publisher = {{Springer}},
  location = {{Berlin ; New York}},
  isbn = {978-3-540-03875-7},
  pagetotal = {482}
}

@article{hunterMatplotlib2DGraphics2007,
  title = {Matplotlib: {{A 2D Graphics Environment}}},
  shorttitle = {Matplotlib},
  author = {Hunter, John D.},
  date = {2007-05},
  journaltitle = {Computing in Science \& Engineering},
  volume = {9},
  number = {3},
  pages = {90--95},
  issn = {1558-366X},
  doi = {10.1109/MCSE.2007.55},
  eventtitle = {Computing in {{Science}} \& {{Engineering}}}
}

@article{jacobsQuantitativeCharacterizationSurface2017,
  title = {Quantitative Characterization of Surface Topography Using Spectral Analysis},
  author = {Jacobs, Tevis D. B. and Junge, Till and Pastewka, Lars},
  date = {2017},
  journaltitle = {Surface Topography: Metrology and Properties},
  shortjournal = {Surf. Topogr.: Metrol. Prop.},
  volume = {5},
  number = {1},
  pages = {013001},
  issn = {2051-672X},
  doi = {10.1088/2051-672X/aa51f8},
  urldate = {2018-06-08},
  langid = {english}
}

@article{jacqDevelopmentThreeDimensionalSemiAnalytical2002,
  title = {Development of a {{Three-Dimensional Semi-Analytical Elastic-Plastic Contact Code}}},
  author = {Jacq, C. and Nélias, D. and Lormand, G. and Girodin, D.},
  date = {2002},
  journaltitle = {Journal of Tribology},
  volume = {124},
  number = {4},
  pages = {653},
  issn = {07424787},
  doi = {10.1115/1.1467920},
  urldate = {2015-12-10},
  langid = {english}
}

@article{johnsonReviewTheoryRolling1966,
  title = {A Review of the Theory of Rolling Contact Stresses},
  author = {Johnson, K. L.},
  date = {1966-01-01},
  journaltitle = {Wear},
  shortjournal = {Wear},
  volume = {9},
  number = {1},
  pages = {4--19},
  issn = {0043-1648},
  doi = {10.1016/0043-1648(66)90010-X},
  urldate = {2023-08-02},
  langid = {english}
}

@article{kapoorSteadyStateSliding1994,
  title = {The Steady State Sliding of Rough Surfaces},
  author = {Kapoor, A. and Williams, J. A. and Johnson, K. L.},
  date = {1994-06-01},
  journaltitle = {Wear},
  shortjournal = {Wear},
  volume = {175},
  number = {1},
  pages = {81--92},
  issn = {0043-1648},
  doi = {10.1016/0043-1648(94)90171-6},
  urldate = {2019-01-14}
}

@article{loSurfaceRougheningContact1999,
  title = {Surface {{Roughening}} and {{Contact Behavior}} in {{Forming}} of {{Aluminum Sheet}}},
  author = {Lo, Sy-Wei and Horng, Tzu-Chern},
  date = {1999-04-01},
  journaltitle = {Journal of Tribology},
  shortjournal = {Journal of Tribology},
  volume = {121},
  number = {2},
  pages = {224--233},
  issn = {0742-4787},
  doi = {10.1115/1.2833925},
  urldate = {2023-08-03}
}

@article{mandelbrotFractalCharacterFracture1984,
  title = {Fractal Character of Fracture Surfaces of Metals},
  author = {Mandelbrot, Benoit B. and Passoja, Dann E. and Paullay, Alvin J.},
  date = {1984-04-19},
  journaltitle = {Nature},
  shortjournal = {Nature},
  volume = {308},
  number = {5961},
  pages = {721--722},
  issn = {0028-0836},
  doi = {10.1038/308721a0},
  urldate = {2016-10-14},
  langid = {english}
}

@article{milaneseEmergenceSelfaffineSurfaces2019,
  title = {Emergence of Self-Affine Surfaces during Adhesive Wear},
  author = {Milanese, Enrico and Brink, Tobias and Aghababaei, Ramin and Molinari, Jean-François},
  date = {2019-03-08},
  journaltitle = {Nature Communications},
  volume = {10},
  number = {1},
  pages = {1116},
  issn = {2041-1723},
  doi = {10.1038/s41467-019-09127-8},
  urldate = {2019-05-21},
  langid = {english}
}

@article{mindlinForcePointInterior1936,
  title = {Force at a {{Point}} in the {{Interior}} of a {{Semi}}‐{{Infinite Solid}}},
  author = {Mindlin, Raymond D.},
  date = {1936-05-01},
  journaltitle = {Journal of Applied Physics},
  volume = {7},
  number = {5},
  pages = {195--202},
  issn = {0021-8979, 1089-7550},
  doi = {10.1063/1.1745385},
  urldate = {2016-10-19}
}

@article{nayakRandomProcessModel1971,
  title = {Random {{Process Model}} of {{Rough Surfaces}}},
  author = {Nayak, P. Ranganath},
  date = {1971-07-01},
  journaltitle = {Journal of Lubrication Technology},
  shortjournal = {J. of Lubrication Tech},
  volume = {93},
  number = {3},
  pages = {398--407},
  issn = {0742-4787},
  doi = {10.1115/1.3451608},
  urldate = {2016-08-29}
}

@article{nohringNonequilibriumPlasticRoughening2022,
  title = {Nonequilibrium Plastic Roughening of Metallic Glasses Yields Self-Affine Topographies with Strain-Rate and Temperature-Dependent Scaling Exponents},
  author = {Nöhring, Wolfram G. and Hinkle, Adam R. and Pastewka, Lars},
  date = {2022-07-21},
  journaltitle = {Physical Review Materials},
  shortjournal = {Phys. Rev. Mater.},
  volume = {6},
  number = {7},
  pages = {075603},
  publisher = {{American Physical Society}},
  doi = {10.1103/PhysRevMaterials.6.075603},
  urldate = {2023-07-12}
}

@article{peiFiniteElementModeling2005,
  title = {Finite Element Modeling of Elasto-Plastic Contact between Rough Surfaces},
  author = {Pei, L. and Hyun, S. and Molinari, J.-F. and Robbins, Mark O.},
  date = {2005-11},
  journaltitle = {Journal of the Mechanics and Physics of Solids},
  shortjournal = {Journal of the Mechanics and Physics of Solids},
  volume = {53},
  number = {11},
  pages = {2385--2409},
  issn = {0022-5096},
  doi = {10.1016/j.jmps.2005.06.008},
  urldate = {2015-12-14}
}

@article{pengMosaicOrganizationDNA1994,
  title = {Mosaic Organization of {{DNA}} Nucleotides},
  author = {Peng, C.-K. and Buldyrev, S. V. and Havlin, S. and Simons, M. and Stanley, H. E. and Goldberger, A. L.},
  date = {1994-02-01},
  journaltitle = {Physical Review E},
  shortjournal = {Phys. Rev. E},
  volume = {49},
  number = {2},
  pages = {1685--1689},
  publisher = {{American Physical Society}},
  doi = {10.1103/PhysRevE.49.1685},
  urldate = {2023-08-03}
}

@article{pengQuantificationScalingExponents1995,
  title = {Quantification of Scaling Exponents and Crossover Phenomena in Nonstationary Heartbeat Time Series},
  author = {Peng, C.‐K. and Havlin, Shlomo and Stanley, H. Eugene and Goldberger, Ary L.},
  date = {1995-03-01},
  journaltitle = {Chaos: An Interdisciplinary Journal of Nonlinear Science},
  shortjournal = {Chaos: An Interdisciplinary Journal of Nonlinear Science},
  volume = {5},
  number = {1},
  pages = {82--87},
  issn = {1054-1500},
  doi = {10.1063/1.166141},
  urldate = {2023-08-03}
}

@article{perssonNatureSurfaceRoughness2005,
  title = {On the Nature of Surface Roughness with Application to Contact Mechanics, Sealing, Rubber Friction and Adhesion},
  author = {Persson, B. N. J. and Albohr, O. and Tartaglino, U. and Volokitin, A. I. and Tosatti, E.},
  date = {2005},
  journaltitle = {Journal of Physics: Condensed Matter},
  shortjournal = {J. Phys.: Condens. Matter},
  volume = {17},
  number = {1},
  pages = {R1},
  issn = {0953-8984},
  doi = {10.1088/0953-8984/17/1/R01},
  urldate = {2016-07-04},
  langid = {english}
}

@book{perssonSlidingFriction2000,
  title = {Sliding {{Friction}}},
  author = {Persson, Bo N. J.},
  editorb = {Von Klitzing, Klaus and Wiesendanger, Roland},
  editorbtype = {redactor},
  date = {2000},
  series = {{{NanoScience}} and {{Technology}}},
  publisher = {{Springer Berlin Heidelberg}},
  location = {{Berlin, Heidelberg}},
  doi = {10.1007/978-3-662-04283-0},
  urldate = {2023-08-22},
  isbn = {978-3-642-08652-6 978-3-662-04283-0}
}

@article{pham-baCreationEvolutionRoughness2021,
  title = {Creation and Evolution of Roughness on Silica under Unlubricated Wear},
  author = {Pham-Ba, Son and Molinari, Jean-François},
  date = {2021-05-15},
  journaltitle = {Wear},
  shortjournal = {Wear},
  volume = {472--473},
  pages = {203648},
  issn = {0043-1648},
  doi = {10.1016/j.wear.2021.203648},
  urldate = {2023-03-27},
  langid = {english}
}

@article{ponsonStatisticalAspectsCrack2016,
  title = {Statistical Aspects in Crack Growth Phenomena: How the Fluctuations Reveal the Failure Mechanisms},
  shorttitle = {Statistical Aspects in Crack Growth Phenomena},
  author = {Ponson, Laurent},
  date = {2016-09-01},
  journaltitle = {International Journal of Fracture},
  shortjournal = {Int J Fract},
  volume = {201},
  number = {1},
  pages = {11--27},
  issn = {1573-2673},
  doi = {10.1007/s10704-016-0117-7},
  urldate = {2019-02-25},
  langid = {english}
}

@article{popovAdhesiveWearParticle2018,
  title = {Adhesive Wear and Particle Emission: {{Numerical}} Approach Based on Asperity-Free Formulation of {{Rabinowicz}} Criterion},
  shorttitle = {Adhesive Wear and Particle Emission},
  author = {Popov, Valentin L. and Pohrt, Roman},
  date = {2018-08-20},
  journaltitle = {Friction},
  shortjournal = {Friction},
  issn = {2223-7704},
  doi = {10.1007/s40544-018-0236-4},
  urldate = {2018-08-23},
  langid = {english}
}

@article{ramisettiAutocorrelationFunctionIsland2011,
  title = {The Autocorrelation Function for Island Areas on Self-Affine Surfaces},
  author = {Ramisetti, Srinivasa B. and Campañá, Carlos and Anciaux, Guillaume and Molinari, Jean-Francois and Müser, Martin H. and Robbins, Mark O.},
  date = {2011},
  journaltitle = {Journal of Physics: Condensed Matter},
  shortjournal = {J. Phys.: Condens. Matter},
  volume = {23},
  number = {21},
  pages = {215004},
  issn = {0953-8984},
  doi = {10.1088/0953-8984/23/21/215004},
  urldate = {2016-07-18},
  langid = {english}
}

@article{renardConstantDimensionalityFault2013,
  title = {Constant Dimensionality of Fault Roughness from the Scale of Micro-Fractures to the Scale of Continents},
  author = {Renard, François and Candela, Thibault and Bouchaud, Elisabeth},
  date = {2013},
  journaltitle = {Geophysical Research Letters},
  volume = {40},
  number = {1},
  pages = {83--87},
  issn = {1944-8007},
  doi = {10.1029/2012GL054143},
  urldate = {2019-09-11},
  langid = {english}
}

@article{rottgerContactEngineeringCreate2022,
  title = {Contact.Engineering—{{Create}}, Analyze and Publish Digital Surface Twins from Topography Measurements across Many Scales},
  author = {Röttger, Michael C. and Sanner, Antoine and Thimons, Luke A. and Junge, Till and Gujrati, Abhijeet and Monti, Joseph M. and Nöhring, Wolfram G. and Jacobs, Tevis D. B. and Pastewka, Lars},
  date = {2022-09},
  journaltitle = {Surface Topography: Metrology and Properties},
  shortjournal = {Surf. Topogr.: Metrol. Prop.},
  volume = {10},
  number = {3},
  pages = {035032},
  publisher = {{IOP Publishing}},
  issn = {2051-672X},
  doi = {10.1088/2051-672X/ac860a},
  urldate = {2023-04-21},
  langid = {english}
}

@article{shvartsTrappedFluidContact2018,
  title = {Trapped Fluid in Contact Interface},
  author = {Shvarts, A. G. and Yastrebov, V. A.},
  date = {2018-10-01},
  journaltitle = {Journal of the Mechanics and Physics of Solids},
  shortjournal = {Journal of the Mechanics and Physics of Solids},
  volume = {119},
  pages = {140--162},
  issn = {0022-5096},
  doi = {10.1016/j.jmps.2018.06.016},
  urldate = {2019-07-08}
}

@article{silvasabinoEvolutionRealContact2023,
  title = {Evolution of the Real Contact Area of Self-Affine Non-{{Gaussian}} Surfaces},
  author = {Silva Sabino, T. and Couto Carneiro, A. M. and Pinto Carvalho, R. and Andrade Pires, F. M.},
  date = {2023-04-15},
  journaltitle = {International Journal of Solids and Structures},
  shortjournal = {International Journal of Solids and Structures},
  volume = {268},
  pages = {112173},
  issn = {0020-7683},
  doi = {10.1016/j.ijsolstr.2023.112173},
  urldate = {2023-04-14},
  langid = {english}
}

@book{simoComputationalInelasticity1998,
  title = {Computational Inelasticity},
  author = {Simo, J. C. and Hughes, Thomas J. R.},
  date = {1998},
  series = {Interdisciplinary Applied Mathematics},
  number = {v. 7},
  publisher = {{Springer}},
  location = {{New York}},
  isbn = {978-0-387-97520-7},
  pagetotal = {392}
}

@article{stanleyFFTBasedMethodRough1997,
  title = {An {{FFT-Based Method}} for {{Rough Surface Contact}}},
  author = {Stanley, H. M. and Kato, T.},
  date = {1997-07-01},
  journaltitle = {Journal of Tribology},
  shortjournal = {J. Tribol},
  volume = {119},
  number = {3},
  pages = {481--485},
  issn = {0742-4787},
  doi = {10.1115/1.2833523},
  urldate = {2015-11-13}
}

@book{taborHardnessMetals1951,
  title = {The Hardness of Metals},
  author = {Tabor, David},
  date = {1951},
  series = {Monographs on the Physics and Chemistry of Materials},
  publisher = {{Clarendon Press}},
  location = {{Oxford}},
  langid = {english}
}

@article{thomsonEffectPlasticDeformation1980,
  title = {The Effect of Plastic Deformation on the Roughening of Free Surfaces of Sheet Metal},
  author = {Thomson, P. F. and Nayak, P. U.},
  date = {1980-01-01},
  journaltitle = {International Journal of Machine Tool Design and Research},
  shortjournal = {International Journal of Machine Tool Design and Research},
  volume = {20},
  number = {1},
  pages = {73--86},
  issn = {0020-7357},
  doi = {10.1016/0020-7357(80)90020-7},
  urldate = {2023-08-03},
  langid = {english}
}

@article{tiwariContactMechanicsSolids2019,
  title = {Contact {{Mechanics}} for {{Solids}} with {{Randomly Rough Surfaces}} and {{Plasticity}}},
  author = {Tiwari, Avinash and Wang, Anle and Müser, Martin H. and Persson, B. N. J.},
  date = {2019-10},
  journaltitle = {Lubricants},
  volume = {7},
  number = {10},
  pages = {90},
  publisher = {{Multidisciplinary Digital Publishing Institute}},
  doi = {10.3390/lubricants7100090},
  urldate = {2020-03-05},
  issue = {10},
  langid = {english}
}

@article{vakisModelingSimulationTribology2018,
  title = {Modeling and Simulation in Tribology across Scales: {{An}} Overview},
  shorttitle = {Modeling and Simulation in Tribology across Scales},
  author = {Vakis, A. I. and Yastrebov, V. A. and Scheibert, J. and Nicola, L. and Dini, D. and Minfray, C. and Almqvist, A. and Paggi, M. and Lee, S. and Limbert, G. and Molinari, J. F. and Anciaux, G. and Aghababaei, R. and Echeverri Restrepo, S. and Papangelo, A. and Cammarata, A. and Nicolini, P. and Putignano, C. and Carbone, G. and Stupkiewicz, S. and Lengiewicz, J. and Costagliola, G. and Bosia, F. and Guarino, R. and Pugno, N. M. and Müser, M. H. and Ciavarella, M.},
  date = {2018-09-01},
  journaltitle = {Tribology International},
  shortjournal = {Tribology International},
  volume = {125},
  pages = {169--199},
  issn = {0301-679X},
  doi = {10.1016/j.triboint.2018.02.005},
  urldate = {2018-10-16}
}

@article{weberMolecularProbesReveal2018,
  title = {Molecular Probes Reveal Deviations from {{Amontons}}’ Law in Multi-Asperity Frictional Contacts},
  author = {Weber, B. and Suhina, T. and Junge, T. and Pastewka, L. and Brouwer, A. M. and Bonn, D.},
  date = {2018-03-01},
  journaltitle = {Nature Communications},
  volume = {9},
  number = {1},
  pages = {888},
  issn = {2041-1723},
  doi = {10.1038/s41467-018-02981-y},
  urldate = {2018-03-12},
  langid = {english}
}

@article{wuSimulationRoughSurfaces2000,
  title = {Simulation of Rough Surfaces with {{FFT}}},
  author = {Wu, Jiunn-Jong},
  date = {2000-01-01},
  journaltitle = {Tribology International},
  shortjournal = {Tribology International},
  volume = {33},
  number = {1},
  pages = {47--58},
  issn = {0301-679X},
  doi = {10.1016/S0301-679X(00)00016-5},
  urldate = {2019-05-20}
}

@article{yastrebovContactRepresentativeRough2012,
  title = {Contact between Representative Rough Surfaces},
  author = {Yastrebov, Vladislav A. and Anciaux, Guillaume and Molinari, Jean-François},
  date = {2012-09-14},
  journaltitle = {Physical Review E},
  volume = {86},
  number = {3},
  issn = {1539-3755, 1550-2376},
  urldate = {2015-11-13},
  langid = {english}
}

\pagebreak
\begin{figure*}
  \centering
  \includegraphics[width=\textwidth]{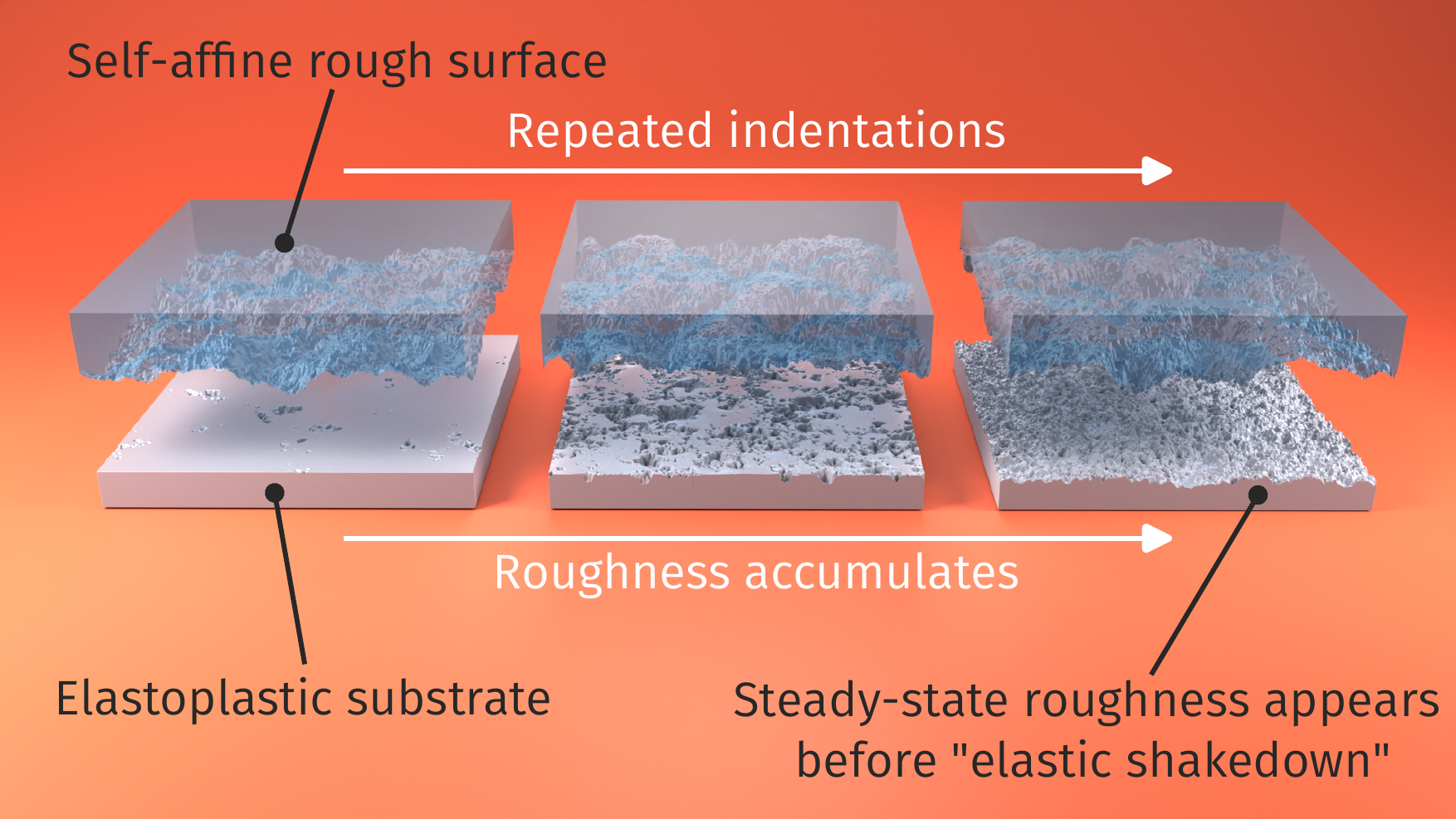}
  \caption{Graphical Abstract}
\end{figure*}
\end{document}